\documentstyle[aps,preprint]{revtex}

\begin{document}
\title{Modeling two-state cooperativity in protein folding}
\author{Ke Fan, Jun Wang, and Wei Wang$^*$}
\address{National Laboratory of Solid State Microstructure 
and Department of Physics, Nanjing University, 
Nanjing 210093, China}
\date{February 19, 2001}
\maketitle
\draft
\begin{abstract}
A protein model with the pairwise interaction energies varying as local environment 
changes, i.e., including some kinds of collective effect between the contacts, is 
proposed. Lattice Monte Carlo simulations on the thermodynamical characteristics and 
free energy profile show a well-defined two-state behavior and cooperativity of folding 
for such a model. As a comparison, related simulations for the usual G\={o} model, where 
the interaction energies are independent of the local conformations, are also made. 
Our results indicate that the evolution of  interactions during the folding process plays 
an important role in the two-state cooperativity in protein folding.
\end{abstract}


Due to the developments of experimental methods and theoretical models, many achievements 
of protein folding have been made recently\cite{dillnsb}. A protein can fold itself to 
its uniquely well-defined native structure in a biologically short time, regardless of 
the huge number of possible conformations, showing a highly cooperatively kinetic behavior. 
It is now clear that the cooperativity of folding may result from the backbone hydrogen 
bonding, sidechain packing and hydrophobic interactions, among them the hydrophobic 
interactions are believed to be the dominant driving force for folding\cite{sorenson}. 
For many small single-domain proteins or lattice proteinlike models, there is a two-state behavior between the unfolded states and the folded native one\cite{jackson,dill95}. 
Recently, Chan and Kaya\cite{Chan} indicated that according to the calorimetric criterion, 
which is widely used in experiments as a condition for two-state folding, popular lattice models, e.g., two-letter HP and 20-letter MJ models, are far from two-state models. 
This may be due to some flawed assumptions in the potential functions used in these models. 
Lattice models usually use statistical potential functions extracted from the pairing frequencies of 20 kinds of amino acids in databases of protein structures\cite{macro85}.
Although these knowledge-based potentials may be a good approximation to the relative 
strength of interactions between the residues in the native state, they provide no 
information about how the interactions evolve during the folding. For computational 
convenience, a common assumption in lattice models is that the interactions are additive, 
and they are the same during the folding as in the native state. This means that the 
interaction energies are conformation-independent. Clearly this is not relevant to the
experimental situation\cite{baldwin}. In fact, as Dill pointed out\cite{dill97}, 
the thermodynamic additivity principle which is widely used in chemistry may be unsuitable 
in biochemistry. Some recent experiments also indicated that the transition state is 
an expanded version of the native state, where the majority of interactions are 
partially formed\cite{fersht}, and their strengths are different from those in the 
native state (with $\Phi<1$). 
That is, these interactions depend on the conformations\cite{baldwin}, especially 
the local structures around the contacts as emphasized recently in 
Ref. \cite{maritan}. 
Previously, the non-additivity was built in a lattice 
model for packing effects\cite{pnas93}; and the hydrophobic force depending on the 
local density of peptide atoms was also taken into account in an off-lattice model
\cite{takada}. Studies on these models show that the introduction of the non-additivity 
is significiant, but the two-state cooperativity of these models is not checked, and 
the effects of the non-additivity on thermodynamics and kinetics of folding need to be 
further studied.

In this paper, we develop a refined G\={o} model where the pairwise interaction energies 
vary as the local environment changes, i.e., some kinds of collective effects between 
contacts are introduced. Our purpose here is to study the two-state cooperativity of 
protein folding and its physical origin with such a model. Our results give a general 
picture about how the conformation-dependent interactions affect the folding kinetics, 
which is consistent with the phenomenological explanation based on experimental results.

We model a polypeptide chain as a self-avoiding chain on a cubic lattice. A contact 
is formed if two residues are space adjacent but not sequence adjacent. If two residues 
form a contact as the same as in the native state, we call this contact a native contact,
otherwise a non-native contact. Following the G\={o} model\cite{go}, only native contacts 
are considered to contribute to the total energy. Different from the G\={o} model, 
we assume that the interaction energies between residues are conformation-dependent, 
and vary with changes of the local environment. To achieve this, we introduce a parameter 
{\sl S} to describe the degree for a residue being ordered relative to the native state. 
For the $i$-th residue in a certain conformation, its degree of order $S_{i}$ is defined as
\begin{equation}
S_{i}=z_{i}/z_{i}^{nat},
\end{equation}
where $z_{i}$ is the number of native contacts in this conformation, $z_{i}^{nat}$ 
is the number of contacts formed in the native state. Obviously, $S_{i}$ varies 
between $0$ (the $i$-th residue being fully disordered) and $1$ (being fully ordered). 
Thus, the 
interaction energy between residues $i$ and $j$, 
$B_{ij}$=-($S_{i}$+$S_{j}$)$\varepsilon$/2 is defined, 
where $S_{i}$ and $S_{j}$ are the degrees of order for residues $i$ and $j$, respectively. 
$\varepsilon$ is the unit of energy and is set to be $1$ in this work. 
The total energy of the 
conformation then is $E=\sum_{i<j}\Delta_{ij}B_{ij}$,
where $\Delta$ is unity when residues $i$ and $j$ form a native contact, and zero otherwise. 
Here, a contact formed between residues $i$ and $j$ may have different energies in different 
conformations, i.e., $B_{ij}$ may change from one conformation to another (for the G\={o} model, 
one always has $B_{ij}=-\varepsilon$). In general, a contact formed between residues $i$ and 
$j$ will stabilize, to some extent, other contacts that residue $i$ or $j$ formed with other 
residues. On the contrary, its breakage may destabilize those contacts as well. Therefore, the 
introduction of the degree of order for a residue into the potential function reflects the 
cooperativity between the residues. Although the correlation distance is small, 
only one lattice unit, the many-body effects are obviously included in our model. 
Figure 1 shows such a collective effect. 
The interaction energies of contact A-B (or B-C) are different when the other 
contact is present or not present. Clearly, the energy of state $I_{3}$ is lower 
than the sum of that of states $I_{1}$ and $I_{2}$, indicating the interaction 
non-additivity. 
Each contact is stabilized by the other contact due to the collective effect. 
Note that in this paper our model 
is called G\={o}+ model to distinguish from the G\={o} model.

Now let us present the Monte Carlo simulations on the thermodynamic and kinetic features for 
both models. The mean first passage time (MFPT), as a common measure of folding rate, is 
calculated by an average of the first passage time (FPT) over 1000 runs. 
Each run begins with a random conformation, and ends when the native state is 
reached for the first time. The FPT is 
the Monte Carlo steps (MCS) consumed in a run.

Generally, as the temperature $T$ decreases, the population of the native state, $P_{N}$, 
increases from zero to about unity. The degree of sharpness of changes in $P_{N}$, similar to 
the ``rapidity" in Ref.\cite{cond}, is a measure of the cooperativity of the folding reaction. 
Figure 2 shows the population $P_{N}$ and the specific heat $C_{v}$ versus 
temperature $T$ for a $36$-mer chain for both models. $P_{N}$ is defined as
$P_{N}=e^{-E_{N}/T}/\sum_{E}\Omega(E) e^{-E/T}$,
where $\Omega(E)$ is the density of states for energy $E$, $E_{N}$ is the energy of the native 
state. $\Omega(E)$ is calculated with the Monte Carlo histogram method\cite{prl89}. 
From Fig.2 
we can see that the folding transition for our G\={o}+ model is much sharper than 
that of the G\={o} model, i.e., a sharper change in $P_{N}$. There is also a single 
peak in $C_{v}$ curve, but it is narrower than that of the G\={o} model. 
For our G\={o}+ model, the maximum of $C_{v}$ 
occurs at a temperature nearly the midpoint temperature of transition with $P_{N}$=1/2, i.e., 
the difference between these two temperatures is quite small. This is consistent with recent 
studies on naturally occuring proteins\cite{cond,prl99}, implying a good cooperativity of 
folding in this model. Differently, such a temperature difference is large for the G\={o} model 
(see Fig.2), indicating that the folding of the G\={o} model is much less cooperative than that 
of the G\={o}+ model. Since the sharpness is only a qualitative description for the transition, 
we further calculate the equilibrium energy distribution at the folding transition temperature, 
$T_{f}$. Figure 3 shows such distributions for both models. Clearly our G\={o}+ model shows a 
good bimodal behavior, and the denatured-state energy is distributed in a narrow region [see 
Fig. 3(a)]. This means clearly a two-state folding and there is basically no intermediate states 
at equilibrium. Differently, for the G\={o} model as shown in Fig. 3(b), there are many 
intermediate states and the bimodal behavior is not so significant as that in Fig.3(a). 
Thus for the G\={o} model the folding is not of a two-state. This is in agreement with 
Chan and Kaya's argument\cite{Chan}.

In experiments, a well-established criterion for two-state folding is that the van't Hoff 
enthalpy $\Delta H_{vH}$ around the transition midpoint is equal, or very close, to the 
calorimetric enthalpy $\Delta H_{cal}$ of the entire transition. In this work, we calculate the 
ratio $\Delta H_{vH}$/$\Delta H_{cal}$ as suggested in Ref.\cite{Chan} (here, the definition of 
$\Delta H_{vH}$/$\Delta H_{cal}$ is equal to $(k_{2})^{2}$ in Ref.\cite{Chan}), and list the 
results in Table I. From Table I, we can clearly see the difference between the G\={o} model and 
our G\={o}+ model. The G\={o} model, which is considered as a model with minimal energetic 
frustrations, does not meet the calorimetric two-state criterion and gives out the value of 
$\Delta H_{vH}$/$\Delta H_{cal}$ far from 1. Nevertheless, our model satisfies the criterion 
quite well (for real proteins, the value of $\Delta H_{vH}$/$\Delta H_{cal}$ is 0.96$\pm 0.03$
\cite{privalov}). This, again, implies the two-state folding and the good cooperativity of our 
G\={o}+ model.

Physically, the high cooperativity of our model may result from the narrow distribution of the 
denatured states and the high population of the native state at the folding temperature (see 
also Figs.2 and 3). 
In our model, the energy spectrum relating to various conformations is 
redistributed, comparing with that of the G\={o} model, due to the collective 
effect between interactions. 
As a result, the energies of non-native conformations are moved to higher energy
levels and a larger energy gap is left between the non-native conformations and 
the native one (for the two models, the energies of the native state are the same). 
The large energy gap makes 
the native state paticularly stable, which is believed to be a necessary condition for 
cooperative folding\cite{nature94}. This may be the physical origin of the two-state 
cooperativity. It can be further explained from the viewpoint of the free energy profile. 
For our G\={o}+ model, as shown in Fig.4, the free energy profiles have broad activation
barriers. 
The broad activation barriers can account for the large movement of transition state 
caused by mutation or temperature changes, and are considered as a common feature of 
the two-state folding\cite{fersht98}. 
Our numerical results are surprisingly consistent with a phenomenological 
speculation for the existence of such a free energy profile in Ref. \cite{fersht98}. 
It should be noted that the broad activation barriers are consistent with the narrow
distribution of the denatured states.

Now let us make a comparison of the foldability based on the plots of the MFPT versus 
$P_{N}$ for both models. Note that we use $P_{N}$ instead of the commonly used temperature 
$T$ in the horizontal axis in Fig.5. 
This is because that an identical condition should be taken for the comparison. 
In lattice simulations, the temperature has an arbitrary unit and also has no direct 
relationship with the real temperature. The comparison between two different models 
at the same temperature may make no sense. 
Nevertheless, at an identical condition with the same $P_{N}$, the differences in 
the foldability can be well-defined. This is similar to other conditions used 
previously\cite{abkevich}. From Fig. 5, we can see that the MFPT for our G\={o}+ model 
shows a slow decrease as $P_{N}$ increases, it reaches a minimum at $P_{N}\approx 0.93$, 
and then it increases. 
For the G\={o} model, there is also a minimum but at $P_{N}\approx 0.71$. 
It is clearly that when the native state is stable (say, $P_{N}\geq$ 0.9), our G\={o}+ 
model folds significantly fast, i.e., the MFPT is smaller with one or two orders of 
magnitude than that of the G\={o} model. 
Physically, this can be explained as follows. From Eq.(1) we can easily see 
that the energy gain of forming a contact is usually smaller for our G\={o}+ model 
than that for the G\={o} model. At high temperatures, entropic contribution is 
dominant to the free energy barrier, and the loss of entropy is always undercompensated 
by the energy gain, thus the G\={o}+ model folds slower for its smaller energy gain. 
Whereas at low temperatures, folding is nearly a downhill process, and the loss of 
entropy is always overcompensated by the energy gain. 
Therefore, for the G\={o}+ model, it is easier to escape from kinetic traps, 
and the folding is faster. Finally, we note that for the two models the pathways 
of reaching the transition state from the denatured state are different. 
Due to the high cooperativity in our G\={o}+ model, a good core, the assembly of 
non-polar residues, is formed much earlier at low temperatures than that in the 
G\={o} model. Detailed kinetic results will be reported elsewhere. 
We also note that similar results are obtained for different chain sizes.

In conclusion, our G\={o}+ model, with many-body interactions depending on the local 
structures included, exhibits a good two-state folding behavior. 
Our results suggest that the evolution of interactions during the folding plays an 
important role in the two-state cooperativity in protein folding. 
We give a possible way how the interactions evolve in the folding, which may capture 
some essential features of the two-state folding. 
We expect further study could provide new insights into the mechanism of protein folding.

We thank H.S. Chan, A. Maritan and D. Thirumalai for useful suggestions. 
This work was supported by the Foundation of NNSF (No.19625409, and No.10074030).

$*$Email address: wangwei@nju.edu.cn

\newpage
\parindent 0pt {\large {\bf Table I:}}
The ratios of $\Delta H_{vH}$ / $\Delta H_{cal}$ for the G\={o} model and 
our G\={o}+ model, respectively. Ten sequences are calculated for each chain size.

\newpage
\parindent 0pt {\large {\bf FIG. 1:}}
Schematic illustration of collective effect between two interactions. 
From a state $I_{0}$ with three unstructured residues, the chain can be settled 
in a state $I_{1}$ (or $I_{2}$) with a contact A-B (or B-C) and an equilibrium 
constant $K_{1}$ (or $K_{2}$). A state $I_{3}$ with two contacts A-B and B-C can 
be reached from state $I_{1}$ or $I_{2}$, but with different equilibrium constants 
$K_{2}\gamma$ or $K_{1}\gamma$. In state $I_{3}$, each interaction is 
stronger by a factor $\gamma$ due to the existence of the other contact.

\parindent 0pt {\large {\bf FIG. 2:}}
Population $P_{N}$ and specific heat $C_{v}$ varying with the temperature $T$ 
for a $36$-mer chain.

\parindent 0pt {\large {\bf FIG. 3:}}
The energy distribution for the same 36-mer used in Fig.2, using 
(a) G\={o}+ potential and 
(b) G\={o} potential at respective folding transition temperature, $T_{f}$. 

\parindent 0pt {\large {\bf FIG. 4:}}
The free energy profile $F(E)=E-TS(E)$ of our G\={o}+ model at different temperatures, 
where entropy $S(E)$ is calculated by using entropy sampling Monte Carlo method\cite{prl93}. Here U, N and TS denote the unfolded state, native state and transition state, 
respectively. Note that the free energy profile at high temperature is overall shifted 
so that the unfolded states are overlapped.

\parindent 0pt {\large {\bf FIG. 5:}}
MFPT versus $P_{N}$ for a 36-mer chain. 


\begin{references}
\bibitem{dillnsb}  
	K.A. Dill and H.S. Chan,
	Nat. Struct. Biol. {\bf 4}, 10, (1997);
	D. Baker,
	Nature (London) {\bf 405}, 39, (2000);
	W.A. Eaton {\sl et al}., 
	Annu. Rev. Biophys. Biomol. Struct. {\bf 29}, 327, (2000);
	J.N. Onuchic, L.-S. Zaida, and P.G. Wolynes,
	Annu. Rev. Phys. Chem. {\bf 48}, 525, (1997);
	C.M. Dobson and M. Karplus,
	Curr. Opin. Struct. Biol. {\bf 9}, 92, (1999);
	J. Wang and W. Wang,
	Nat. Struct. Biol. {\bf 6}, 1033, (1999).
\bibitem{sorenson}  
	J.M. Sorenson and T. Head-Gordon,
	Fold. Des. {\bf 3}, 523, (1998).
\bibitem{jackson}  
	S.E. Jackson,
	Fold. Des. {\bf 3}, R81, (1998).
\bibitem{dill95}  
	H.S. Chan, S. Bromberg, and K.A. Dill, 
	Phil. Trans. R. Soc. Lond. B {\bf 348}, 61, (1995);
	D.K. Klimov and D. Thirumalai, 
	Fold. Des. {\bf 3}, 127, (1998);
\bibitem{Chan}  
	H.S. Chan, 
	Proteins {\bf 40}, 543, (2000);
	H. Kaya and H.S. Chan, 
	Proteins {\bf 40}, 637, (2000);
	H. Kaya and H.S. Chan, 
	Phys. Rev. Lett. {\bf 85}, 4823, (2000);
\bibitem{macro85}  
	S. Miyazawa and R.L. Jernigan, 
	Macromolecules {\bf 18}, 534, (1985);
	A. Koliski, A. Godzik, and J. Skolnick, 
	J. Chem. Phys. {\bf 98}, 7420, (1993).
\bibitem{baldwin}  
	A.R. Fersht {\sl et al}., 
	Proc. Natl. Acad. Sci. USA {\bf 91}, 10426, (1994);
	R.L. Baldwin, 
	Nature (London) {\bf 369}, 183, (1994).
\bibitem{dill97}  
	K.A. Dill, 
	J. Biol. Chem. {\bf 272}, 701, (1997).
\bibitem{fersht}  
	D.E. Otzen {\sl et al}., 
	Proc. Natl. Acad. Sci. USA {\bf 91}, 10422, (1994);
	F. Chiti {\sl et al}., 
	Nat. Struct. Biol. {\bf 6}, 1005, (1999);
	J.C. Martinez and L. Serrano, 
	{\sl ibid.} {\bf 6}, 1010, (1999);
	D.S. Riddle {\sl et al}., 
	{\sl ibid.} {\bf 6}, 1016, (1999).
\bibitem{maritan}  
	J.R. Banavar and A. Maritan, 
	Proteins {\bf 42}, 433, (2001).
\bibitem{pnas93}  
	C.J. Camacho and D. Thirumalai, 
	Proc. Natl. Acad. Sci. USA {\bf 90}, 6369, (1993).
\bibitem{takada}  
	S. Takada, Z. Luthey-Schulten, and P.G. Wolynes, 
	J. Chem. Phys. {\bf 110}, 11616, (1999).
\bibitem{go}  
	N. G\={o}, 
	Annu. Rev. Biophys. Bioeng. {\bf 12}, 183, (1983).
\bibitem{cond}  
	F. Cecconi {\sl et al}., 
	cond-matt/0101229.
\bibitem{prl89}  
	A.M. Ferrenberg and R.H. Swendsen,
	Phys. Rev. Lett. {\bf 63}, 1195, (1989);
	N.D. Socci and J.N. Onuchic,
	J. Chem. Phys. {\bf 103}, 4732, (1995).
\bibitem{prl99}  
	C. Micheletti {\sl et al}., 
	Phys. Rev. Lett. {\bf 82}, 3372, (1999).
\bibitem{privalov}  
	P.L. Privalov, 
	Adv. Protein Chem. {\bf 33}, 167, (1979).
\bibitem{nature94}  
	A. Sali, E.I. Shakhnovich, and M. Karplus, 
	Nature (London) {\bf 369}, 248, (1994).
\bibitem{fersht98}  
	M. Oliveberg {\sl et al}.,
	J. Mol. Biol. {\bf 277}, 933, (1998);
	D.E. Otzen {\sl et al}.,
	Biochemistry {\bf 38}, 6499, (1999);
	M. Oliveberg, 
	Acc. Chem. Res. {\bf 31}, 765, (1998).
\bibitem{abkevich}  
	V.I. Abkevich, A.V. Gutin, and E.I. Shakhnovich,
	J. Mol. Biol. {\bf 252}, 460, (1995);
	D.K. Klimov and D. Thirumalai, 
	Proteins {\bf 26}, 411, (1996).
\bibitem{prl93}  
	J. Lee, 
	Phys. Rev. Lett. {\bf 71}, 211, (1993);
	M.H. Hao and H.A. Scheraga, 
	J. Phys. Chem. {\bf 98}, 4940, (1994).


\end{references}
\end{document}